%% file: PRL_main.tex
\documentclass[
superscriptaddress,
 amsmath,amssymb,
 aps,
prl,
twocolumn,
]{revtex4-2}
\newcommand{\thor}{$^{229}$Th}

\newcommand{\lisaf}{LiSrAlF$_6$}

\usepackage[dvipsnames]{xcolor}

\usepackage{braket}
\usepackage{graphicx}
\usepackage[colorlinks=true, allcolors=blue]{hyperref}
\usepackage{gensymb}

\usepackage{upgreek}

\usepackage{float}  
\usepackage{xcolor}

\definecolor{ricky}{cmyk}{0, 0.7808, 0.4429, 0.1412}

\usepackage{braket}
\usepackage{graphicx}

\usepackage[colorlinks=true, allcolors=blue]{hyperref}
\usepackage[version=3]{mhchem}

\usepackage{upgreek}
\usepackage{tikz-cd}
\tikzcdset{arrow style=tikz, diagrams={>=stealth}}

\begin{document}

\title{Photo-induced Quenching of the $^{229}$Th Isomer in a Solid-State Host}

\author{J. E. S. Terhune}
\affiliation{Department of Physics and Astronomy, University of California, Los Angeles, CA 90095, USA}
\author{R. Elwell}
\affiliation{Department of Physics and Astronomy, University of California, Los Angeles, CA 90095, USA}
\author{H. B. Tran Tan}
\affiliation{Department of Physics, University of Nevada, Reno, Nevada 89557, USA}
\affiliation{Los Alamos National Laboratory, P.O. Box 1663, Los Alamos, New Mexico 87545, USA} 
\author{U. C. Perera}
\affiliation{Department of Physics, University of Nevada, Reno, Nevada 89557, USA}
\author{H. W. T. Morgan}
\affiliation{Department of Chemistry and Biochemistry, University of California, Los Angeles, Los Angeles, CA 90095, USA}
\author{A. N. Alexandrova}
\affiliation{Department of Chemistry and Biochemistry, University of California, Los Angeles, Los Angeles, CA 90095, USA}
\author{Andrei Derevianko}
\affiliation{Department of Physics, University of Nevada, Reno, Nevada 89557, USA}
\author{Eric R. Hudson}
\affiliation{Department of Physics and Astronomy, University of California, Los Angeles, CA 90095, USA}
\affiliation{Challenge Institute for Quantum Computation, University of California Los Angeles, Los Angeles, CA, USA}
\affiliation{Center for Quantum Science and Engineering, University of California Los Angeles, Los Angeles, CA, USA}
\date{\today} 

\begin{abstract}
The population dynamics of the \thor{} isomeric state is studied in a solid-state host under laser illumination. 
A photoquenching process is observed, where off-resonant vacuum-ultraviolet (VUV) radiation leads to relaxation of the isomeric state. 
The cross-section for this photoquenching process is measured and a model for the decay process, where photoexcitation of electronic states within the material bandgap opens an internal conversion decay channel, is presented and appears to reproduce the measured cross-section. 
\end{abstract}

\maketitle

The \thor{} nuclear isomeric state provides a laser-accessible, nuclear transition that can be driven even when doped into a high-bandgap solid~\cite{Hudson2008, Rellergert2010}. 
The unique properties of the system are expected to allow access to a number of novel applications, including the construction of a solid-state nuclear clock~\cite{Rellergert2010}.
With the recent laser excitation and measurement of the isomer energy and lifetime in these systems~\cite{Tiedau2024, Elwell2024, Zhang2024-Th229Comb, zhang2024thf}, work can now begin towards constructing and optimizing timekeeping devices based on thorium-doped crystals. 

The stability of an optical clock operating by collecting fluorescence from the clock transition is limited to~\cite{MartinBoydThesis2007}

    $\sigma = \sqrt{(T_e + T_c)/T}/(2\pi Q S)$,

where $Q=f_0/\Delta f$ is the transition quality factor, $f_0$ is the transition frequency, $\Delta f$ is the transition linewidth, $S$ is the signal-to-noise ratio (SNR), $T_e$ is the excitation time, $T_c$ is the fluorescence collection time, and $T$ is the averaging time. 
Thus, optimization of clock performance requires maximizing $S$, while minimizing $T_e + T_c$ without degrading $Q$.
The transition linewidth in \thor{} doped fluoride crystals is expected to be $\sim 1$~kHz~\cite{Rellergert2010}, which is $10^6\times$ larger than the intrinsic linewidth of the transition. 
As a result, simply put, optimum clock performance is realized by exciting all \thor{} nuclei in a time of order 1~ms and having them decay as quickly as possible. 

Given that laser excitation of a nucleus in a solid-state environment has only just been demonstrated~\cite{Tiedau2024, Elwell2024}, it is not yet known what values of $S$, $T_e$, and $T_c$ are attainable. 
In particular, Ref.~\cite{Elwell2024} observed that not all \thor{} in the crystal were laser excitable and speculated that impurities and/or crystal defects could quench the transition or alter its decay lifetime. 
This observation suggests that control of these electronic defect states could perhaps be used to optimize clock performance by improving $S$ and potentially reducing $T_e$ and $T_c$.

To better understand these processes, here we present a study of the excitation dynamics of \thor{} nuclei doped into a \lisaf{} crystal. 
We find that the vacuum-ultraviolet (VUV) photons used to excite the \thor{} nuclei also lead to a process that quenches nuclear excitation.
We study this process to determine the cross-section of the photo-quenching process and provide a description of  mechanisms that could be the origin of the phenomena.
We conclude with a discussion of the use of photoquenching for optimizing solid-state nuclear clock performance and suggest directions of future work to better understand the process. 

\begin{figure}
    \centering
    \includegraphics[width=\linewidth]{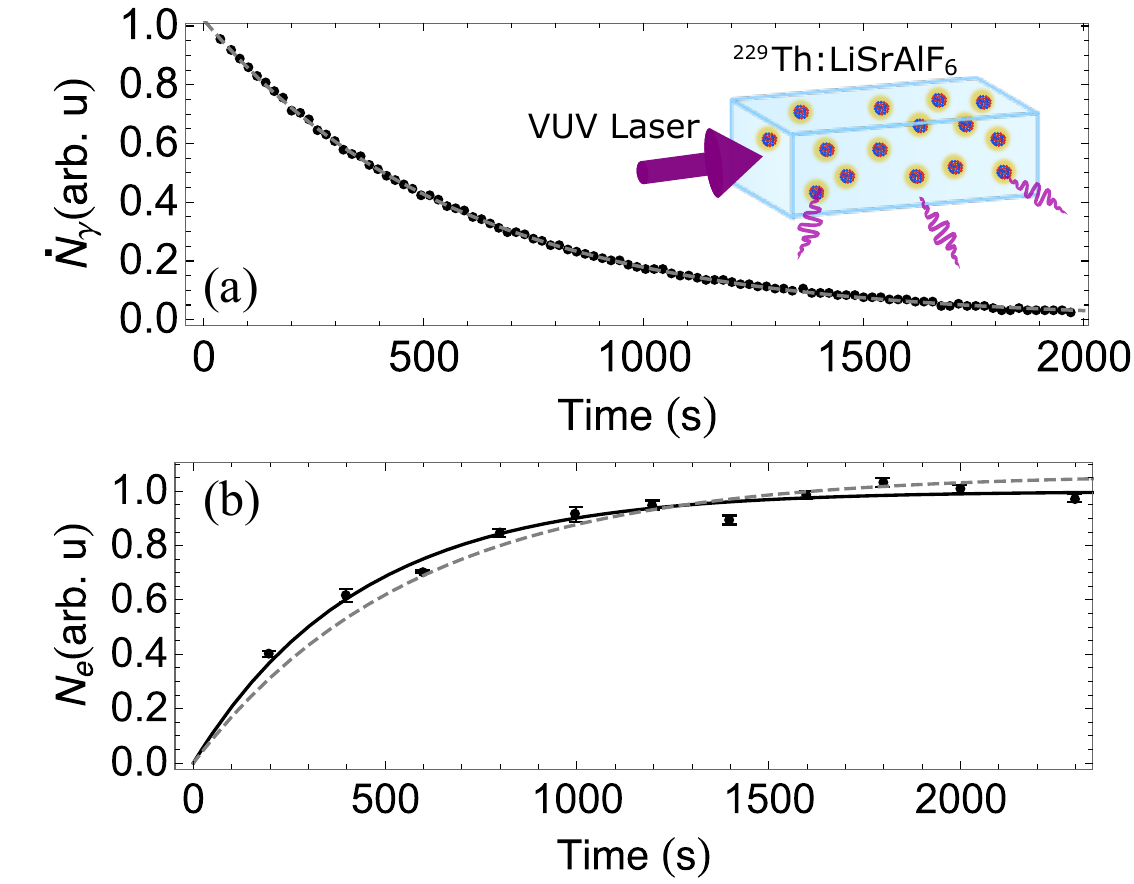}
    \caption{(a) Photon emission rate of $^{229m}$Th following laser excitation determining the radiative decay rate in \lisaf{} to be $\tau = 573.4(29)$~s. (b) The total number of excited \thor{} as a function of laser excitation time. The excited state population grows with a lifetime of $\tau_{\textrm{eff}} = 431(70)$~s. The dashed gray line is a fit to the data with $\tau_{\textrm{eff}}$ fixed to $\tau$.}
    \label{fig:DecayPlusExcitation}
\end{figure}

The experiments reported here were conducted via monitoring the resulting fluorescence from a \thor{}:\lisaf{} crystal (crystal 3.1 from Ref.~\cite{Elwell2024}) following illumination with a VUV laser (see inset Fig.~\ref{fig:DecayPlusExcitation}). 
Using the system described in Ref.~\cite{JeetThesis2018, Elwell2024}, VUV radiation was produced via resonance-enhanced four-wave mixing of two pulsed dye lasers (PDL) in Xe gas. 
The frequency of the first PDL (referred to as the UV PDL), $\omega_u$, was locked to the $5p^{6 ~1}S_0~\rightarrow~5p^5\left(^2P^{\circ}_{3/2} \right) 6p~^2\left[1/2\right]_0$ two-photon transition of Xe at $\sim$ 249.63 nm. 
The frequency of the second PDL, $\omega_v$, was scanned to produce VUV radiation in the Xe cell given by the difference mixing relation $\omega = 2\omega_{u} - \omega_v$. 
All three laser beams then impinge off-axis with respect to a MgF$_2$ lens, whose chromatic dispersion is used with downstream pinholes to spatially filter the VUV beam and pass it towards the \thor{}:\lisaf{} crystal.
The laser system delivers 30 pulses per second to the crystal with a typical VUV pulse energy of $\sim 5~\mu$J/pulse.

The crystal under study is held in a vacuum chamber consisting of a crystal mount, two VUV-sensitive photomultiplier tubes (PMTs; Hamamatsu R6835), and a pneumatic shutter system to protect the PMTs from direct exposure to the VUV laser.
The PMTs are operated in a cathode-grounded configuration, and their output waveforms recorded by a 1~Gs/s waveform digitizer (CAEN DT5751) for subsequent post-processing. 
The VUV laser beam terminates on a custom VUV energy detector.
The crystal chamber is maintained with an Ar atmosphere at a pressure of $\sim$10$^{-2}$~mbar to provide high VUV transmission while minimizing browning of optics due to hydrocarbon deposition~\cite{JeetThesis2018, Elwell2024}.

We model the laser excitation and quenching of the \thor{} nuclei with a simple rate equation. 
Assuming the quenching process is linear in photon number, the number of excited thorium nuclei is:
\begin{align}
    \dot{N}_e = -\frac{N_e}{\tau} + \sigma_e \dot{\phi}\left(p N_g - \frac{g_g}{g_e} N_e\right) - \sigma_q \dot{\phi} N_e,
    \label{eq:rate_eqn}
\end{align}
where $N_g$ is the number of ground state nuclei, $\tau$ is the isomeric state radiative lifetime in the crystal,  $g_g$ ($g_e$) is the degeneracy of the ground (excited) state, $\sigma_e$ is the wavelength-dependent absorption cross section, $\dot{\phi}$ is the number of photons per second per unit area passing through the crystal, and $\sigma_q$ is the cross-section describing the strength of the photo-quenching.
The coefficient $p$, which we dub the participation factor, is introduced to account for the observation in Ref.~\cite{Elwell2024} that not all \thor{} nuclei present in the crystal participate in the nuclear transition. 

In order to quantify the quenching it is necessary to first measure the other parameters of the model. 
By exciting the nuclear transition for $T_e = 1200$~s and then collecting the resulting fluorescence for $T_c = 2000$~s (see Fig.~\ref{fig:DecayPlusExcitation}(a)), $\tau$ is found from a nonlinear least-squares fit to be $\tau = 573.4(29)$~s -- here and throughout () denote a 68\% confidence interval.
It is difficult to independently measure $p$ as the number of excited nuclei is also affected by the quenching process. 
Nonetheless, as seen below, the quenching process is not detectable at lower laser power.
Therefore, using data taken at low laser power, the radiative lifetime extracted from Fig.~\ref{fig:DecayPlusExcitation}(a), the system detection efficiency, and the solution of Eq.~\eqref{eq:rate_eqn}, it is estimated that $p = 0.14(1)$. This yields an effective density of $n_p = 7.0(5) \times 10^{14}$~cm$^{-3}$ of participating \thor{} nuclei. 

To ascertain the relative importance of the quenching dynamics, we first monitor the number of excited thorium nuclei as a function of excitation time. 
Given the narrow linewidth of the nuclear transition ($\sim 1$~kHz) relative to the excitation laser linewidth ($\sim 10$~GHz), $\sigma_e\dot{\phi} $ is on the order of $ 10^{-9}$. 
Likewise, the number of excited \thor{} nuclei as a function of laser excitation time can be approximated as $N_e(t) \approx \sigma_e p \dot{\phi}\tau (1-\exp{\left(-t_e(1/\tau + \sigma_q \dot{\phi)}\right)}) = \sigma_e p \dot{\phi}\tau(1-\exp{(-t_e/\tau_{\textrm{eff}})})$.
Figure~\ref{fig:DecayPlusExcitation}(b), shows the total number of detected photons $N_\gamma^{(q)}$ following excitation, a quantity proportional to $N_e$, with a VUV laser power of $P \approx 90$~$\mu$W versus laser exposure time $t_e$.
The solid black line is a variable lifetime fit of $N_e(t)$ to the data.
This fit finds an effective lifetime of $\tau_{eff} = 431(70)$~s, while the dashed gray line is a fit of $N_e(t)$ with $\tau_{eff}$ fixed to $\tau$.
Both a reduced $\chi^2$ and Akaike information criteria model test prefer the model with variable lifetime, though neither is conclusive.
Further, the fitted effective lifetime of $\tau_{\textrm{eff}} = 431(70)$~s differs from $\tau$ by approximately two standard errors and suggests quenching by an amount of $\sigma_q \approx 1$~Mb.

\begin{figure}
    \centering
    \includegraphics[width=\linewidth]{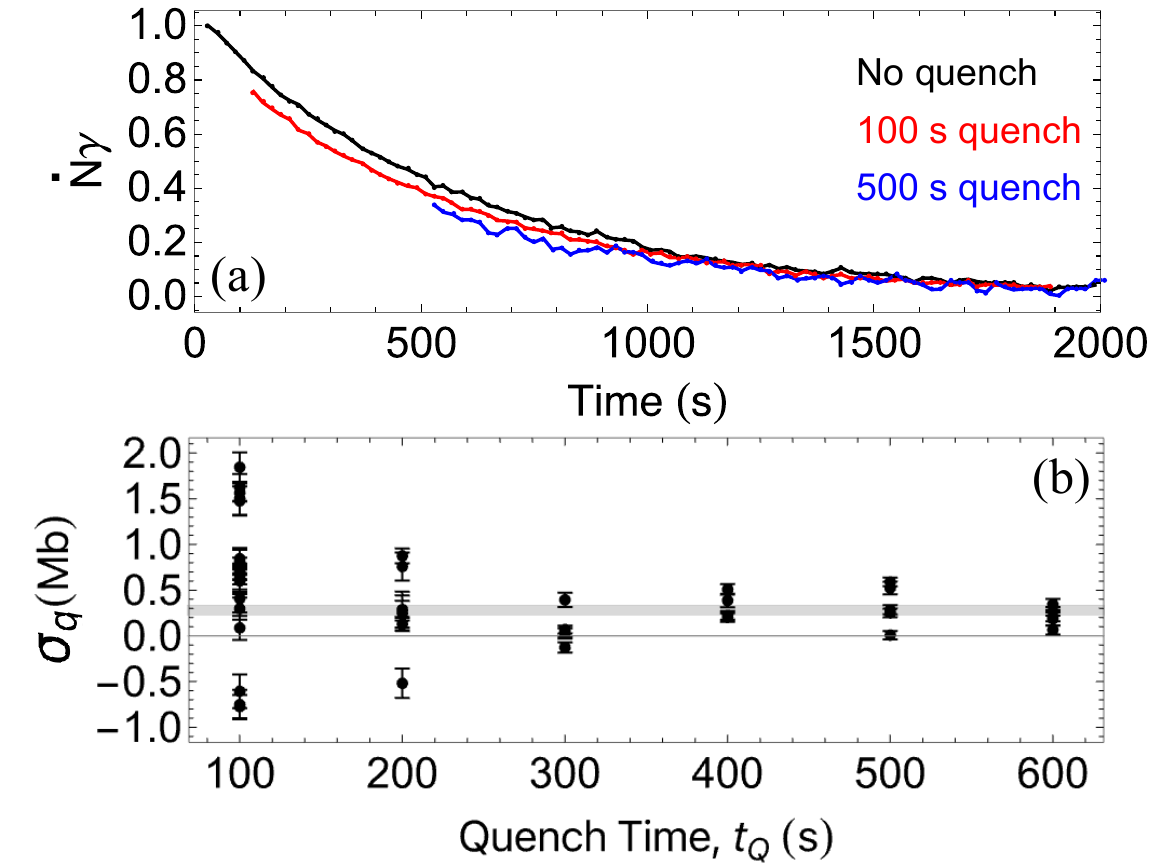}
    \caption{(a) Photon emission rate of $^{229m}$Th following laser excitation and a variable amount of quenching by an off-resonant laser. Each trace represents the average of several trials and clearly show the reduced $^{229m}$Th population with quenching. (b) The extracted quenching cross-section $\sigma_q$ from each trial is shown as a function of quenching time $t_q$. At short quenching times, variability in laser parameters leads to a spread in excitation probability and therefore the extracted $\sigma_q$. }
    \label{fig:TimeQuench}
\end{figure}

To confirm the quenching process, we perform an additional experiment where after laser excitation for $T_e = 1200$~s, the excitation laser is detuned by approximately 100~GHz so that it does not efficiently excite the nuclei, but may still quench excited nuclei. 
We then record the emitted photons as a function of this variable `quench time' $t_q$, as shown in Fig.~\ref{fig:TimeQuench}(a) for $t_q = 0$~s, $100$~s, and $500$~s.
As can be seen, the total number of emitted photons decreases with increase quench time. 
This process is repeated for several quench times, and $\sigma_q$ is extracted from the data with a model that accounts for the small off-resonant excitation rate that can occur during the quench time; larger laser detunings are not possible in our system.
The results are shown in Fig.~\ref{fig:TimeQuench}(b) and lead to an average value of 
$\sigma_q = 0.29(3)_{stat}(12)_{sys} \,\mathrm{Mb}$, where the systematic error is dominated by uncertainty in the VUV beam waist.

\begin{figure}
    \includegraphics[width=\linewidth]{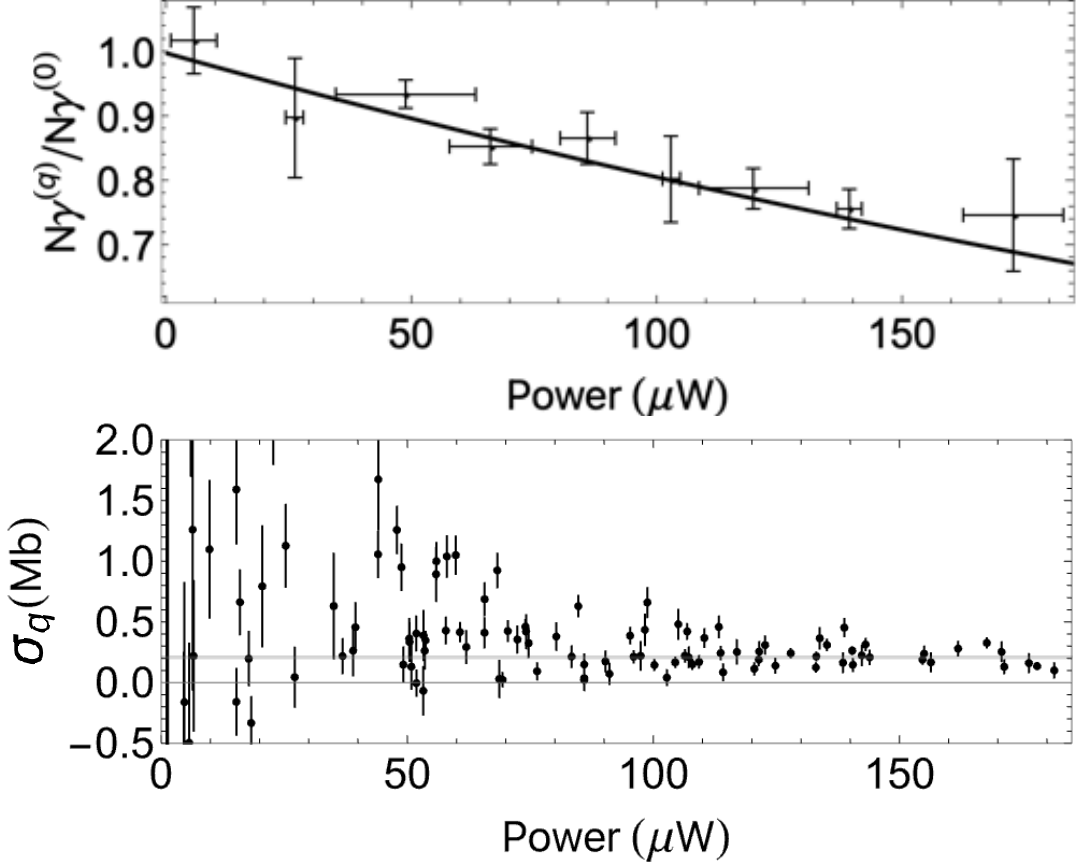}
    \caption{(a) Ratio of quenched to unquenched photo emission  following laser excitation at variable quenching laser power. Each data point is an average of at least 5 trials.(b) The extracted quenching cross-section from each trial as a function of quenching laser power. The weighted average of these points is  $\sigma_q = 0.23(1)_{stat}(9)_{sys} \, \mathrm{Mb}$}
 \label{fig:PowerQuench}
\end{figure}

To further investigate the photoquenching process, the dependence of $\sigma_q$ on the photon flux $\dot{\phi}$ was studied.
The data in Fig.~\ref{fig:PowerQuench} is taken in the same manner as that of Fig.~\ref{fig:TimeQuench}, but here the quenching laser is applied for a fixed time of $t_q = 500$~s and its power varied. 
The laser power is controlled during the quench process by detuning the UV PDL a variable amount away from the Xe resonance. 
In this way, the change in Xe excitation controls the realized VUV power. 
Shown alongside the data as a black line is a prediction of the quenching based on the fit to the data in Fig.~\ref{fig:TimeQuench}, which was taken at 90~$\mu$ W, and an assumption of linear dependence of the photo-quenching on $\dot{\phi}$. These data do not support a $\dot{\phi}^2$ dependence. A $\dot{\phi}^2$ dependence would result in the solid black prediction line curving down and away from the data points at higher laser power since the degree of quenching would increase quadratically with increasing laser power.

While this prediction roughly agrees with the data, it is possible that the quenching process saturates at around 100~$\mu$W. 
Confirmation of this quenching is not possible in the present apparatus due to laser instabilities above 150~$\mu$W. 
Saturation of the photoquenching could arise, for example, if not all of the excited \thor{} undergo the photo-quenching process, which as described later could arise if the quenching process depends on the nature of localized electronic states within the band gap. 
A fit to the data that allows for a variable quenching fraction indicates that roughly 50\% of the \thor{} can be photoquenched and experience a photoquenching cross section of $\sigma_q \approx 4.6 \pm 1.1 Mb$.
The possibility of quenching saturation will be the study of future work when reliable higher power laser operation is available.

Given that the next nuclear excited state is nearly 30~keV above the isomer state, the quenching process is presumably driven by photoexcitation of crystal electronic states.
Such spatially-localized, electronic defect states with energies inside the band gap can result from impurities, crystalline defects, and doping \thor{} into \lisaf{}.
Ref.~\cite{morgan_internal_conversion_2024} has demonstrated that if the energy, $\varepsilon_d$, of the defect state $\ket{d}$ is less than the nuclear excitation energy, $\varepsilon_n$, and the minimum of the valence band $\varepsilon_m$ is such that $\varepsilon_d - \varepsilon_m \ge \varepsilon_n$, the nuclear excitation can relax by resonant excitation of a defect-hole pair. 
This process likely occurs in $\sim 10^{-3}$~s and is presumably responsible for the deviation of the participation fraction $p$ from unity. 
It also implies that the \thor{} responsible for the signal observed here are near defect states that have energies that are either greater than the isomer energy ($\varepsilon_d > \varepsilon_n$) or so low that there is no valence band state available for resonant excitation to the defect state ($\varepsilon_d < \varepsilon_n + \varepsilon_m$).

To excite defect states with $\varepsilon_d > \varepsilon_n$, the energy of the VUV laser used here requires either (i) a multi-photon or (ii) single photon + multi-phonon absorption process. 
Since the data in Fig.~\ref{fig:PowerQuench}(a) do not show a $\dot{\phi}^2$ dependence, the second possibility is more likely.
Phonons can be generated via thermal excitation or optical absorption by defects throughout the crystal when lattice relaxation is involved~\cite{Huang1981}. 
Excitation of defects with energy $\varepsilon_d < \varepsilon_n + \varepsilon_m$ can be accomplished by single-photon absorption from the quenching laser. 
Ref.~\cite{Elwell2024} has calculated that for \thor{}:\lisaf{} $\varepsilon_n + \varepsilon_m \approx 5$~eV.
Analysis of the photoluminescence of the \thor{}:\lisaf{} crystals when illuminated by the VUV laser reveals a roughly 100~nm wide feature centered around 425~nm (see SI), indicative of defect states satisfying $\varepsilon_d < 5$~eV.

\begin{figure}[t!]
    \centering
    \includegraphics[width=\linewidth]{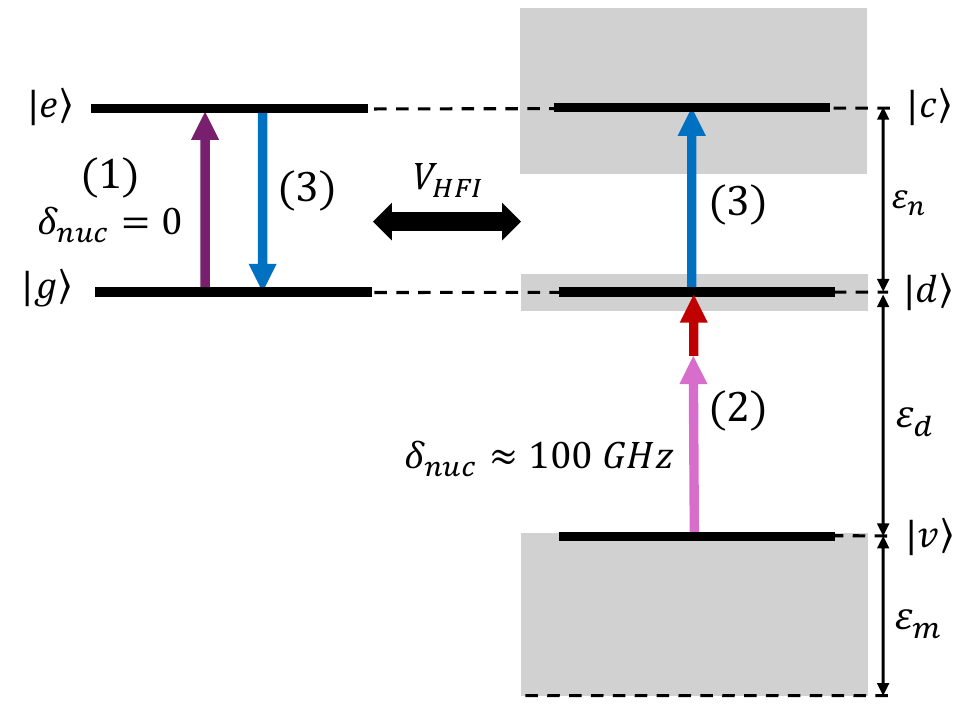}
    \caption{Example photo-quenching mechanism. Valence, defect, and conduction band states are denoted by ($v,d,c$), respectively, while the \thor{} ground and isomeric states are denoted by ($g,e$), respectively. (1) Resonant VUV laser drives the nucleus to the isomeric state. (2) The VUV laser and, in one process, multi-phonon absorption drives an electronic transition from $\ket{v}$ to $\ket{d}$. 
    (3) If $\varepsilon_d +\varepsilon_n > \varepsilon_c$, an internal conversion channel is opened and the isomer can decay by exciting an electron into the conduction band.}
    \label{fig:readout_scheme}
\end{figure}

Regardless of the mechanism, if $\ket{d}$ is populated and $\varepsilon_d + \varepsilon_n \geq \varepsilon_c$, where $\varepsilon_c$ is the energy of the bottom of the conduction band, the isomer can undergo an internal conversion (IC) decay process by resonantly exciting the defect state electron to the conduction band (see Fig.~\ref{fig:readout_scheme}). 
Following similar arguments to those in Ref.~\cite{morgan_internal_conversion_2024}, this IC rate is much faster than the radiative nuclear decay rate. 
This defect to conduction band IC process may be responsible for the observed photoquenching.

To estimate the rate of this IC process, we perform a more comprehensive rate equation analysis that involves the relevant nuclear and electronic degrees of freedom (see SI). 
The resulting photoquenching cross-section is found as $\sigma_q \approx \lambda^2/(2\pi)(\Gamma_\mathrm{IC}/\Gamma_\mathrm{L})$, where $\Gamma_\mathrm{IC}$ is the internal conversion rate, $\Gamma_\mathrm{L} = 2\pi\times 15$~GHz is the bandwidth of the laser system~\cite{Elwell2024}, and $\lambda$ is the wavelength of the nuclear transition.
Thus, the observed value of $\sigma_q \approx 0.25\, \mathrm{Mb}$ translates to an IC rate of $\Gamma_\mathrm{IC} \approx 2\pi \times 100\,\mathrm{s}^{-1}$.
While internal conversion rates from defect states into the conduction have not been calculated, this value of $\Gamma_{\textrm{IC}}$ is of the same order of magnitude as the partial IC rate for $5f$ states in neutral Th~\cite{Bilous_2018} and as shown in Ref.~\cite{morgan_internal_conversion_2024} the electronic defects associated with \thor{} doping into \lisaf{} have significant $5f$ character. Estimates in solid-state hosts~\cite{morgan_internal_conversion_2024} indicate a similar IC rate. 

One immediate consequence of this process is that it produces electronic population in the conduction band. 
As a result, this process could be confirmed by the observation of photons whose energy is on the order of the \lisaf{} band gap ($\sim$ 10 eV) or the electron itself. 
Such detection schemes may be beneficial in improving $S$ for \thor{} clock operation.
Further, if the location of the defect states can be determined, perhaps strong resonant excitation of the defects can be used strongly enhance the photoquenching rate. 
This raises the possibility of improved performance for the solid-state nuclear clock since the photoquenching provides  control of the nuclear excited state lifetime. 
For example, pulsed excitation to the defect states with a strong laser could be used to quickly quench the excited nuclear state population, which could be detected via conduction band electrons or their $\sim 10$~eV fluorescence. 
This would obviate the need to wait for the $\sim 1000$~s required for the nucleus to decay during the clock interrogation protocol, translating to improvement in clock instability $\sigma$. 

In summary, photoquenching of the \thor{} isomer when doped in a solid-state host is observed and the cross-section is measured.
The process appears to result from an internal conversion decay channel that is opened by photoexcitation of electronic defects.
A simple model suggests an IC decay rate via theses defects of $\sim 2\pi\times 100$~s$^{-1}$, which is similar to recent estimates for IC decay via $5f$ electrons~\cite{Bilous_2018,morgan_internal_conversion_2024}.
This effect may be quite useful in optimizing the performance of a future solid-state clock as it could reduce the clock interrogation time and increase the clock signal to noise ratio. 

\section{Acknowledgment}
This work was supported by NSF awards PHYS-2013011 and PHY-2207546, and ARO award W911NF-11-1-0369.
ERH acknowledges institutional support by the NSF QLCI Award OMA-2016245.
This work used Bridges-2 at Pittsburgh Supercomputing Center through allocation PHY230110 from the Advanced Cyberinfrastructure Coordination Ecosystem: Services \& Support (ACCESS) program, which is supported by National Science Foundation grants \#2138259, \#2138286, \#2138307, \#2137603, and \#2138296.

\bibliographystyle{apsrev4-2}
\bibliography{ref}

\newpage

\appendix
\include{PRL_SI.tex}

\end{document}

%% file: PRL_SI.tex
\section{Supplemental Information}
\section{Defect State Based Quenching}
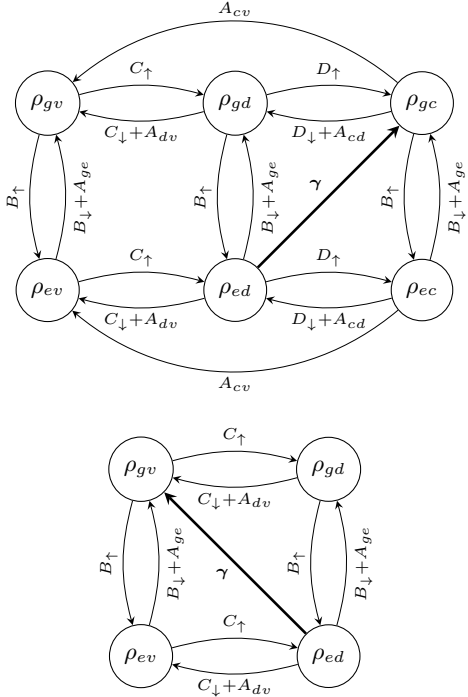
\begin{figure}[h!]
\centering
    \input{diagram_1}
    \input{diagram_2}
\caption{Directed graphs describing the dynamics of populations in the combined nuclear and isomeric system. The top diagram represents all possible couplings between the populations. The bottom represents the simplification where one neglects population in the conduction band.}
\label{fig:directed_graph}
\end{figure}

The observed quenching could be the result of an internal conversion process whereby the \thor{} nucleus gives its energy to a defect state electron, promoting it to the conduction band. The dynamics are described by a series of rate equations that are represented graphically in Fig.~\ref{fig:directed_graph}.

In our analysis, we analyze the populations in the system represented by $\rho_{(g,e)(v,d,c)}$, where $(g,e)$ represent the nuclear ground and excited state, and $(v,d,c)$ represents the nearby electron in the valence band, defect state, and conduction band, respectively. All of the $A$-coefficients represent spontaneous emission rates from the various states. $B_{\uparrow}$ represents the photo-absorption cross-section $\sigma$ discussed in the text multiplied by the photon flux $\Dot{\phi}$. $B_{\downarrow}$ is the same, but multiplied by the appropriate degeneracy factors. The coefficient $C_{\uparrow,\downarrow}$ represent the photon + multi-phonon absorption / emission cross sections for the valence-to-defect transitions, and $D_{\uparrow,\downarrow}$ represent photo-absorption/emission cross-sections for the defect-to-conduction transition. Unlike the $B$-coefficients, $C_{\uparrow,\downarrow}$ and $D_{\uparrow,\downarrow}$ are assumed to be very broad. The $\gamma$ term represents the rate of internal conversion of the defect state electron to the conduction band.

We make the simplifying approximation that the conduction band states are essentially never populated, with $\rho_{gc}$ essentially populated only to immediately relax into $\rho_{gv}$. As such, we can simplify into the approximate system depicted in the bottom of Fig.~\ref{fig:directed_graph}. From this graph we can read out the rate equations for the excited state populations as
\begin{equation}
\begin{aligned}
    \Dot{\rho}_{ev} = &B_{\uparrow}\rho_{gv} - \left( A_{ge} + B_{\downarrow} \right) \rho_{ev} \\&- C_{\uparrow} \rho_{ev} + \left( C_{\downarrow} + A_{dv} \right) \rho_{ed},\\
    \Dot{\rho}_{ed} = &B_{\uparrow}\rho_{gd} - \left( A_{ge} + B_{\downarrow} \right) \rho_{ed} \\ &+ C_{\uparrow} \rho_{ev} - \left( C_{\downarrow} + A_{dv} + \gamma \right) \rho_{ed}.
\end{aligned}
\end{equation} Assuming the defect state lifetime is of the order of a few nanoseconds, reasonable values of $C_{\uparrow,\downarrow}$ are $\sim 100$~Hz. From numerical simulation, it can be shown that the terms $B_{\uparrow}\rho_{gd}$ and $\left( A_{ge} + B_{\downarrow} \right) \rho_{ed}$ in $\Dot{\rho}_{ed}$ can be neglected, and that $\Dot{\rho}_{ed}$ rapidly relaxes to zero. As such, we obtain $\rho_{ed} \approx C_{\uparrow}\rho_{ev}/\left( C_{\downarrow} + A_{dv} + \gamma \right)$. We can now substitute this back into our expression for $\Dot{\rho}_{ev}$ to obtain
\begin{equation}
\begin{aligned}
    \Dot{\rho}_{ev} = &B_{\uparrow}\rho_{gv} - \left( A_{ge} + B_{\downarrow} \right) \rho_{ev} \\&- C_{\uparrow} \rho_{ev} + \frac{\left( C_{\downarrow} + A_{dv} \right)C_{\uparrow}\rho_{ev}} {\left( C_{\downarrow} + A_{dv} + \gamma \right)} \\
    = &B_{\uparrow}\rho_{gv} - \left( A_{ge} + B_{\downarrow} \right) \rho_{ev} - \frac{C_{\uparrow} \gamma}{\left( C_{\downarrow} + A_{dv} + \gamma \right)} \rho_{ev}.
\end{aligned}
\end{equation}
Noting that the observed number of excited nuclei is given by $N_e= \rho_{ev}N_0$, $N_g \approx \rho_{gv}N_0$, $A_{ge}=1/\tau$, $B_{\uparrow} = p\sigma \Dot{\phi}$, $B_{\downarrow}=\frac{g_g}{g_e}\sigma\Dot{\phi}$, $C_{\uparrow} = C \Dot{\phi}$, and assuming $A_{dv} \gg \gamma, C_{\downarrow}$, we obtain
\begin{equation}
    \Dot{N_e} =  -\frac{N_e}{\tau} + \sigma \dot{\phi}\left(p N_g - \frac{g_g}{g_e} N_e\right) - \frac{C \Dot{\phi} \gamma}{ A_{dv} } N_e,
\end{equation} 
where $C$ is a constant proportional to the photo-absorption cross-section corresponding to the electronic valence band to defect state transition, $A_{dv} \sim$ 1 ns is the rate of decay from the defect state back to the valence band, and $\gamma$ is the rate of the IC process. We thus see that a quenching term can arise that is proportional to the photon flux, and we can identify $C\gamma/A_{dv}$ with the quenching cross-section $\alpha_q$. Assuming $C \approx \lambda^2/(2\pi)(A_{dv}/\Gamma_L) $, where $\Gamma_L \approx$ 15 GHz is the bandwidth of the laser system~\cite{Elwell2024}, then we find that $\gamma \approx$ 400 Hz matches the experimentally determined $\sigma_q$. While internal conversion rates from the $5f$ defect state into the conduction have not been calculated, this value is consistent with both the estimates of Ref.~\cite{morgan_internal_conversion_2024} and the partial IC rate calculated for $5f$ states in neutral Th~\cite{Bilous_2018}.

\section{Photoluminescence Spectrum of thorium-doped crystals}
\begin{figure}[t!]
    \centering
    \includegraphics[width=245pt]{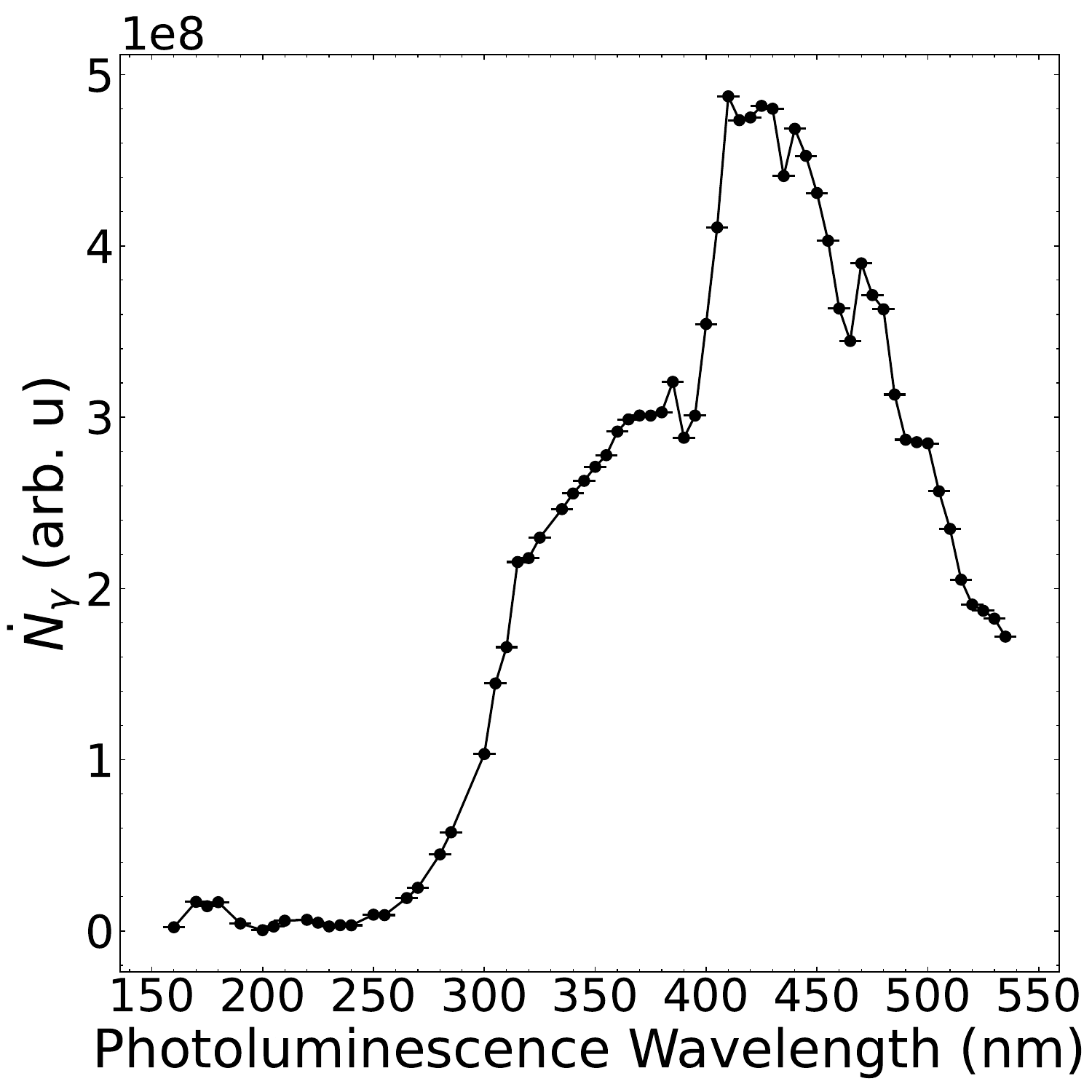}
    \caption{Photoluminescence spectrum from a \thor{}:\lisaf{} crystal when illuminated by a 148.9 nm laser. The horizontal error bars represent the manufacturer provided spectral width of the light that exits that monochromator and is incident on the PMT.}
    \label{fig:photoluminescence_spectrum}
\end{figure}

To probe for the presence of defect states with energy $\varepsilon_d < \varepsilon_n$, the photoluminescence spectrum of \thor{}:\lisaf{} is measured (see Fig. ~\ref{fig:photoluminescence_spectrum}). 
This spectrum is obtained by continuously illuminating the crystal with a 148.9 nm laser and collecting the emitted photoluminescence in a Seya-Namioka VUV monochromator (McPherson 203). 
The count rate of photons exiting the monochromator was obtained using two different PMTs, the Hammamatsu R6836 and H12386-210, in order to scan from roughly 160~nm to 535~nm.

Fig. ~\ref{fig:photoluminescence_spectrum}, reveals a spectrum that peaks around 425 nm. 
This fluorescence could result in several scenarios that all generically result in defect states with $\varepsilon_d < \varepsilon_n + \varepsilon_m$.
For example, the VUV laser could resonantly excite electrons several eV below the top of the valence band to a defect state $\gtrsim 3$ eV above the valence band. 
Subsequent particle-hole recombination in the valence band, as well as non-radiative decays occurring in the defect band, would lead to the observed spectrum and provide states with 
This process would also create phonons, which may allow the aforementioned single photon + multi-phonon excitation into defect states that have energy $\epsilon_d > \epsilon_n$.

Thus, this spectrum appears to support the mechanisms posited here to explain the quenching.
Further work is, of course, needed to confirm these explanations and differentiate the relative contributions of the different mechanisms. 
For example, measurement of $\sigma_q$ as a function of temperature would likely change the rate of the phonon-enabled process.

%% file: diagram_1.tex
\[
\begin{tikzcd}[row sep=5em, column sep=5em, cells={nodes={draw, circle}}]
    \rho_{gv} \arrow[r, bend left=15, "C_{\uparrow}"] \arrow[swap, d, bend right=15, "B_{\uparrow}"{rotate=90,anchor=south,yshift=0}]  & \rho_{gd} \arrow[swap, d, bend right=15, "B_{\uparrow}"{rotate=90,anchor=south,yshift=0}] \arrow[l, bend left=15, "C_{\downarrow} + A_{dv}"] \arrow[r, bend left=15, "D_{\uparrow}"]  & \arrow[swap, ll, bend right=40, "A_{cv}"] \rho_{gc} \arrow[swap, d, bend right=15, "B_{\uparrow}"{rotate=90,anchor=south,yshift=0}] \arrow[l, bend left=15, "D_{\downarrow} + A_{cd}"] \\
    \rho_{ev} \arrow[r, bend left=15, "C_{\uparrow}"] \arrow[swap, u, bend right=15, "B_{\downarrow} + A_{ge}"{rotate=90,anchor=south,yshift=-12}] & \rho_{ed} \arrow[l, bend left=15, "C_{\downarrow} + A_{dv}"] \arrow[swap, u,  bend right=15,"B_{\downarrow} + A_{ge}" {rotate=90,anchor=south,yshift=-12}] \arrow[r, bend left=15, "D_{\uparrow}"] \arrow[ru, line width=1, "\boldsymbol{\gamma}"]& \rho_{ec} \arrow[swap, u,  bend right=15,"B_{\downarrow} + A_{ge}" {rotate=90,anchor=south,yshift=-12}] \arrow[l, bend left=15, "D_{\downarrow} + A_{cd}"] \arrow[ll, bend left=40, "A_{cv}"]
\end{tikzcd}
\]

%% file: diagram_2.tex
\[
\begin{tikzcd}[row sep=5em, column sep=5em, cells={nodes={draw, circle}}]
    \rho_{gv} \arrow[r, bend left=15, "C_{\uparrow}"] \arrow[swap, d, bend right=15, "B_{\uparrow}"{rotate=90,anchor=south,yshift=0}]  & \rho_{gd} \arrow[swap, d, bend right=15, "B_{\uparrow}"{rotate=90,anchor=south,yshift=0}] \arrow[l, bend left=15, "C_{\downarrow} + A_{dv}"] \\
    \rho_{ev} \arrow[r, bend left=15, "C_{\uparrow}"] \arrow[swap, u, bend right=15, "B_{\downarrow} + A_{ge}"{rotate=90,anchor=south,yshift=-12}] & \rho_{ed} \arrow[l, bend left=15, "C_{\downarrow} + A_{dv}"] \arrow[swap, u,  bend right=15,"B_{\downarrow} + A_{ge}" {rotate=90,anchor=south,yshift=-12}]  \arrow[lu, line width=1, "\boldsymbol{\gamma}"]
\end{tikzcd}
\]

%% file: PRL_main.bbl
\begin{thebibliography}{11}%
\makeatletter
\providecommand \@ifxundefined [1]{%
 \@ifx{#1\undefined}
}%
\providecommand \@ifnum [1]{%
 \ifnum #1\expandafter \@firstoftwo
 \else \expandafter \@secondoftwo
 \fi
}%
\providecommand \@ifx [1]{%
 \ifx #1\expandafter \@firstoftwo
 \else \expandafter \@secondoftwo
 \fi
}%
\providecommand \natexlab [1]{#1}%
\providecommand \enquote  [1]{``#1''}%
\providecommand \bibnamefont  [1]{#1}%
\providecommand \bibfnamefont [1]{#1}%
\providecommand \citenamefont [1]{#1}%
\providecommand \href@noop [0]{\@secondoftwo}%
\providecommand \href [0]{\begingroup \@sanitize@url \@href}%
\providecommand \@href[1]{\@@startlink{#1}\@@href}%
\providecommand \@@href[1]{\endgroup#1\@@endlink}%
\providecommand \@sanitize@url [0]{\catcode `\\12\catcode `\$12\catcode `\&12\catcode `\#12\catcode `\^12\catcode `\_12\catcode `\%12\relax}%
\providecommand \@@startlink[1]{}%
\providecommand \@@endlink[0]{}%
\providecommand \url  [0]{\begingroup\@sanitize@url \@url }%
\providecommand \@url [1]{\endgroup\@href {#1}{\urlprefix }}%
\providecommand \urlprefix  [0]{URL }%
\providecommand \Eprint [0]{\href }%
\providecommand \doibase [0]{https://doi.org/}%
\providecommand \selectlanguage [0]{\@gobble}%
\providecommand \bibinfo  [0]{\@secondoftwo}%
\providecommand \bibfield  [0]{\@secondoftwo}%
\providecommand \translation [1]{[#1]}%
\providecommand \BibitemOpen [0]{}%
\providecommand \bibitemStop [0]{}%
\providecommand \bibitemNoStop [0]{.\EOS\space}%
\providecommand \EOS [0]{\spacefactor3000\relax}%
\providecommand \BibitemShut  [1]{\csname bibitem#1\endcsname}%
\let\auto@bib@innerbib\@empty
\bibitem [{\citenamefont {Hudson}\ \emph {et~al.}(2008)\citenamefont {Hudson}, \citenamefont {Vutha}, \citenamefont {Lamoreaux},\ and\ \citenamefont {DeMille}}]{Hudson2008}%
  \BibitemOpen
  \bibfield  {author} {\bibinfo {author} {\bibfnamefont {E.~R.}\ \bibnamefont {Hudson}}, \bibinfo {author} {\bibfnamefont {A.~C.}\ \bibnamefont {Vutha}}, \bibinfo {author} {\bibfnamefont {S.~K.}\ \bibnamefont {Lamoreaux}},\ and\ \bibinfo {author} {\bibfnamefont {D.}~\bibnamefont {DeMille}},\ }\href@noop {} {\bibfield  {journal} {\bibinfo  {journal} {Int. Conf. Atm. Phys., poster M028}\ } (\bibinfo {year} {2008})}\BibitemShut {NoStop}%
\bibitem [{\citenamefont {Rellergert}\ \emph {et~al.}(2010)\citenamefont {Rellergert}, \citenamefont {DeMille}, \citenamefont {Greco}, \citenamefont {Hehlen}, \citenamefont {Torgerson},\ and\ \citenamefont {Hudson}}]{Rellergert2010}%
  \BibitemOpen
  \bibfield  {author} {\bibinfo {author} {\bibfnamefont {W.~G.}\ \bibnamefont {Rellergert}}, \bibinfo {author} {\bibfnamefont {D.}~\bibnamefont {DeMille}}, \bibinfo {author} {\bibfnamefont {R.~R.}\ \bibnamefont {Greco}}, \bibinfo {author} {\bibfnamefont {M.~P.}\ \bibnamefont {Hehlen}}, \bibinfo {author} {\bibfnamefont {J.~R.}\ \bibnamefont {Torgerson}},\ and\ \bibinfo {author} {\bibfnamefont {E.~R.}\ \bibnamefont {Hudson}},\ }\href {https://doi.org/10.1103/PhysRevLett.104.200802} {\bibfield  {journal} {\bibinfo  {journal} {Phys. Rev. Lett.}\ }\textbf {\bibinfo {volume} {104}},\ \bibinfo {pages} {200802} (\bibinfo {year} {2010})}\BibitemShut {NoStop}%
\bibitem [{\citenamefont {Tiedau}\ \emph {et~al.}(2024)\citenamefont {Tiedau}, \citenamefont {Okhapkin}, \citenamefont {Zhang}, \citenamefont {Thielking}, \citenamefont {Zitzer}, \citenamefont {Peik}, \citenamefont {Schaden}, \citenamefont {Pronebner}, \citenamefont {Morawetz}, \citenamefont {Toscani De~Col} \emph {et~al.}}]{Tiedau2024}%
  \BibitemOpen
  \bibfield  {author} {\bibinfo {author} {\bibfnamefont {J.}~\bibnamefont {Tiedau}}, \bibinfo {author} {\bibfnamefont {M.~V.}\ \bibnamefont {Okhapkin}}, \bibinfo {author} {\bibfnamefont {K.}~\bibnamefont {Zhang}}, \bibinfo {author} {\bibfnamefont {J.}~\bibnamefont {Thielking}}, \bibinfo {author} {\bibfnamefont {G.}~\bibnamefont {Zitzer}}, \bibinfo {author} {\bibfnamefont {E.}~\bibnamefont {Peik}}, \bibinfo {author} {\bibfnamefont {F.}~\bibnamefont {Schaden}}, \bibinfo {author} {\bibfnamefont {T.}~\bibnamefont {Pronebner}}, \bibinfo {author} {\bibfnamefont {I.}~\bibnamefont {Morawetz}}, \bibinfo {author} {\bibfnamefont {L.}~\bibnamefont {Toscani De~Col}}, \emph {et~al.},\ }\href {https://journals.aps.org/prl/accepted/2c07aYbeC981d47c171619f5604116053962ac79a} {\bibfield  {journal} {\bibinfo  {journal} {Phys. Rev. Lett.}\ }\textbf {\bibinfo {volume} {132}},\ \bibinfo {pages} {182501} (\bibinfo {year} {2024})}\BibitemShut {NoStop}%
\bibitem [{\citenamefont {Elwell}\ \emph {et~al.}(2024)\citenamefont {Elwell}, \citenamefont {Schneider}, \citenamefont {Jeet}, \citenamefont {Terhune}, \citenamefont {Morgan}, \citenamefont {Alexandrova}, \citenamefont {Tran~Tan}, \citenamefont {Derevianko},\ and\ \citenamefont {Hudson}}]{Elwell2024}%
  \BibitemOpen
  \bibfield  {author} {\bibinfo {author} {\bibfnamefont {R.}~\bibnamefont {Elwell}}, \bibinfo {author} {\bibfnamefont {C.}~\bibnamefont {Schneider}}, \bibinfo {author} {\bibfnamefont {J.}~\bibnamefont {Jeet}}, \bibinfo {author} {\bibfnamefont {J.}~\bibnamefont {Terhune}}, \bibinfo {author} {\bibfnamefont {H.}~\bibnamefont {Morgan}}, \bibinfo {author} {\bibfnamefont {A.}~\bibnamefont {Alexandrova}}, \bibinfo {author} {\bibfnamefont {H.}~\bibnamefont {Tran~Tan}}, \bibinfo {author} {\bibfnamefont {A.}~\bibnamefont {Derevianko}},\ and\ \bibinfo {author} {\bibfnamefont {E.~R.}\ \bibnamefont {Hudson}},\ }\href@noop {} {\  (\bibinfo {year} {2024})}\BibitemShut {NoStop}%
\bibitem [{\citenamefont {Zhang}\ \emph {et~al.}(2024{\natexlab{a}})\citenamefont {Zhang}, \citenamefont {Ooi}, \citenamefont {Higgins}, \citenamefont {Doyle}, \citenamefont {von~der Wense}, \citenamefont {Beeks}, \citenamefont {Leitner}, \citenamefont {Kazakov}, \citenamefont {Li}, \citenamefont {Thirolf}, \citenamefont {Schumm},\ and\ \citenamefont {Ye}}]{Zhang2024-Th229Comb}%
  \BibitemOpen
  \bibfield  {author} {\bibinfo {author} {\bibfnamefont {C.}~\bibnamefont {Zhang}}, \bibinfo {author} {\bibfnamefont {T.}~\bibnamefont {Ooi}}, \bibinfo {author} {\bibfnamefont {J.~S.}\ \bibnamefont {Higgins}}, \bibinfo {author} {\bibfnamefont {J.~F.}\ \bibnamefont {Doyle}}, \bibinfo {author} {\bibfnamefont {L.}~\bibnamefont {von~der Wense}}, \bibinfo {author} {\bibfnamefont {K.}~\bibnamefont {Beeks}}, \bibinfo {author} {\bibfnamefont {A.}~\bibnamefont {Leitner}}, \bibinfo {author} {\bibfnamefont {G.~A.}\ \bibnamefont {Kazakov}}, \bibinfo {author} {\bibfnamefont {P.}~\bibnamefont {Li}}, \bibinfo {author} {\bibfnamefont {P.~G.}\ \bibnamefont {Thirolf}}, \bibinfo {author} {\bibfnamefont {T.}~\bibnamefont {Schumm}},\ and\ \bibinfo {author} {\bibfnamefont {J.}~\bibnamefont {Ye}},\ }\href {https://doi.org/10.1038/s41586-024-07839-6} {\bibfield  {journal} {\bibinfo  {journal} {Nature}\ }\textbf {\bibinfo {volume} {633}},\ \bibinfo {pages} {63} (\bibinfo {year} {2024}{\natexlab{a}})}\BibitemShut {NoStop}%
\bibitem [{\citenamefont {Zhang}\ \emph {et~al.}(2024{\natexlab{b}})\citenamefont {Zhang}, \citenamefont {von~der Wense}, \citenamefont {Doyle}, \citenamefont {Higgins}, \citenamefont {Ooi}, \citenamefont {Friebel}, \citenamefont {Ye}, \citenamefont {Elwell}, \citenamefont {Terhune}, \citenamefont {Morgan}, \citenamefont {Alexandrova}, \citenamefont {Tan}, \citenamefont {Derevianko},\ and\ \citenamefont {Hudson}}]{zhang2024thf}%
  \BibitemOpen
  \bibfield  {author} {\bibinfo {author} {\bibfnamefont {C.}~\bibnamefont {Zhang}}, \bibinfo {author} {\bibfnamefont {L.}~\bibnamefont {von~der Wense}}, \bibinfo {author} {\bibfnamefont {J.~F.}\ \bibnamefont {Doyle}}, \bibinfo {author} {\bibfnamefont {J.~S.}\ \bibnamefont {Higgins}}, \bibinfo {author} {\bibfnamefont {T.}~\bibnamefont {Ooi}}, \bibinfo {author} {\bibfnamefont {H.~U.}\ \bibnamefont {Friebel}}, \bibinfo {author} {\bibfnamefont {J.}~\bibnamefont {Ye}}, \bibinfo {author} {\bibfnamefont {R.}~\bibnamefont {Elwell}}, \bibinfo {author} {\bibfnamefont {J.~E.~S.}\ \bibnamefont {Terhune}}, \bibinfo {author} {\bibfnamefont {H.~W.~T.}\ \bibnamefont {Morgan}}, \bibinfo {author} {\bibfnamefont {A.~N.}\ \bibnamefont {Alexandrova}}, \bibinfo {author} {\bibfnamefont {H.~B.~T.}\ \bibnamefont {Tan}}, \bibinfo {author} {\bibfnamefont {A.}~\bibnamefont {Derevianko}},\ and\ \bibinfo {author} {\bibfnamefont {E.~R.}\ \bibnamefont {Hudson}},\ }\href {http://arxiv.org/abs/2410.01753} {\bibfield  {journal} {\bibinfo
  {journal} {Nature (in press)}\ } (\bibinfo {year} {2024}{\natexlab{b}})},\ \Eprint {https://arxiv.org/abs/2410.01753} {arXiv:2410.01753} \BibitemShut {NoStop}%
\bibitem [{\citenamefont {Boyd}(2007)}]{MartinBoydThesis2007}%
  \BibitemOpen
  \bibfield  {author} {\bibinfo {author} {\bibfnamefont {M.~M.}\ \bibnamefont {Boyd}},\ }\emph {\bibinfo {title} {High Precision Spectroscopy of Strontium in an Optical Lattice: Towards a New Standard for Frequency and Time}},\ \href {https://jila.colorado.edu/sites/default/files/2019-05/boyd_thesis.pdf} {Ph.D. thesis},\ \bibinfo  {school} {University of Colorado, Boulder} (\bibinfo {year} {2007})\BibitemShut {NoStop}%
\bibitem [{\citenamefont {Jeet}(2018)}]{JeetThesis2018}%
  \BibitemOpen
  \bibfield  {author} {\bibinfo {author} {\bibfnamefont {J.}~\bibnamefont {Jeet}},\ }\emph {\bibinfo {title} {Search for the low lying transition in the $^{229}$Th Nucleus}},\ \href {https://escholarship.org/uc/item/8wk771ch} {Ph.D. thesis},\ \bibinfo  {school} {University of California, Los Angeles} (\bibinfo {year} {2018})\BibitemShut {NoStop}%
\bibitem [{\citenamefont {Morgan}\ \emph {et~al.}(2024)\citenamefont {Morgan}, \citenamefont {{Tran Tan}}, \citenamefont {Elwell}, \citenamefont {Alexandrova}, \citenamefont {Hudson}, \citenamefont {Derevianko}, \citenamefont {Tan}, \citenamefont {Elwell}, \citenamefont {Alexandrova}, \citenamefont {Hudson},\ and\ \citenamefont {Derevianko}}]{morgan_internal_conversion_2024}%
  \BibitemOpen
  \bibfield  {author} {\bibinfo {author} {\bibfnamefont {H.~W.~T.}\ \bibnamefont {Morgan}}, \bibinfo {author} {\bibfnamefont {H.~B.}\ \bibnamefont {{Tran Tan}}}, \bibinfo {author} {\bibfnamefont {R.}~\bibnamefont {Elwell}}, \bibinfo {author} {\bibfnamefont {A.~N.}\ \bibnamefont {Alexandrova}}, \bibinfo {author} {\bibfnamefont {E.~R.}\ \bibnamefont {Hudson}}, \bibinfo {author} {\bibfnamefont {A.}~\bibnamefont {Derevianko}}, \bibinfo {author} {\bibfnamefont {H.~B.~T.}\ \bibnamefont {Tan}}, \bibinfo {author} {\bibfnamefont {R.}~\bibnamefont {Elwell}}, \bibinfo {author} {\bibfnamefont {A.~N.}\ \bibnamefont {Alexandrova}}, \bibinfo {author} {\bibfnamefont {E.~R.}\ \bibnamefont {Hudson}},\ and\ \bibinfo {author} {\bibfnamefont {A.}~\bibnamefont {Derevianko}},\ }\bibfield  {journal} {\bibinfo  {journal} {Phys. Rev. Lett. (under review)}\ }\href {https://doi.org/10.48550/arXiv.2411.15641} {10.48550/arXiv.2411.15641} (\bibinfo {year} {2024}),\ \Eprint {https://arxiv.org/abs/2411.15641} {arXiv:2411.15641
  [physics.atom-ph]} \BibitemShut {NoStop}%
\bibitem [{\citenamefont {Huang}(1981)}]{Huang1981}%
  \BibitemOpen
  \bibfield  {author} {\bibinfo {author} {\bibfnamefont {K.}~\bibnamefont {Huang}},\ }\href {https://doi.org/10.1080/00107518108231558} {\bibfield  {journal} {\bibinfo  {journal} {Contemporary Physics}\ }\textbf {\bibinfo {volume} {22}},\ \bibinfo {pages} {599} (\bibinfo {year} {1981})}\BibitemShut {NoStop}%
\bibitem [{\citenamefont {Bilous}\ \emph {et~al.}(2018)\citenamefont {Bilous}, \citenamefont {Minkov},\ and\ \citenamefont {P\'alffy}}]{Bilous_2018}%
  \BibitemOpen
  \bibfield  {author} {\bibinfo {author} {\bibfnamefont {P.~V.}\ \bibnamefont {Bilous}}, \bibinfo {author} {\bibfnamefont {N.}~\bibnamefont {Minkov}},\ and\ \bibinfo {author} {\bibfnamefont {A.}~\bibnamefont {P\'alffy}},\ }\href {https://doi.org/10.1103/PhysRevC.97.044320} {\bibfield  {journal} {\bibinfo  {journal} {Phys. Rev. C}\ }\textbf {\bibinfo {volume} {97}},\ \bibinfo {pages} {044320} (\bibinfo {year} {2018})}\BibitemShut {NoStop}%
\end{thebibliography}%
